\documentclass[12pt]{article}
\usepackage{amsthm}
\usepackage{graphicx}
\usepackage{amsmath}
\usepackage{amssymb}
\usepackage{amsfonts}
\usepackage{slashed}
\usepackage{afterpage}
\usepackage{cite}
\usepackage{color}
\definecolor{verde}{cmyk}{.83,.21,1,.08}

\def\nn{\nonumber}

\def\be{\begin{equation}}
\def\ee{\end{equation}}
\def\bea{\begin{eqnarray}}
\def\eea{\end{eqnarray}}

\newcommand{\del}{\partial}

\newcommand{\eqn}[1]{(\ref{#1})}

\newcommand{\no}{\nonumber \\}
\newcommand{\la}{\label}

\begin{document}

\begin{titlepage}
\begin{flushright}
ICCUB-13-244
\end{flushright}

\begin{center}

\baselineskip=24pt

{\Large\bf High energy bosons do not propagate}

\baselineskip=14pt

\vspace{1cm}

{M. A.\ Kurkov$^{1,2}$, Fedele Lizzi$^{1,2,3}$ and Dmitri Vassilevich$^{4}$}
\\[6mm]
$^1${\it Dipartimento di Fisica, Universit\`{a} di Napoli
{\sl Federico II}}
\\[4mm]
$^2${\it INFN, Sezione di Napoli}
\\[4mm]
$^3$ {\it Departament de Estructura i Constituents de la Mat\`eria,
\\Institut de Ci\'encies del Cosmos,
Universitat de Barcelona,\\
Barcelona, Catalonia, Spain}
\\[4mm]
$^4$ {\it CMCC-Universidade Federal do ABC, Santo Andr\'e, S.P., Brazil
}

\bigskip

{\small\tt
Kurkov@na.infn.it, fedele.lizzi@na.infn.it, dvassil@gmail.com}

\end{center}

\vskip 2 cm

\begin{abstract}
We discuss the propagation of bosons (scalars, gauge fields and gravitons) at high energy in the context of the spectral action. Using heat kernel techniques, we find that in the high-momentum limit the quadratic part of the action does not contain positive powers of the derivatives. We interpret this as the fact that the two point Green functions vanish for nearby points, where the proximity scale is given by the inverse of the cutoff.
\end{abstract}

\end{titlepage}

There are indications that at very high energy, of the order of Planck mass $m_p=10^{19}$~GeV the behaviour of particles is profoundly altered by the onset of gravitational effects. The first to notice this has been Bronstein~\cite{Bronstein} in 1936 and since then there have been several attempts to describe the quantum field theory of fields at high energy or small distances. Also in string theory the very high energy behaviour in the scattering of particles~\cite{AmatiCiafaloniVeneziano,GrossMende} shows the existence of some sort of generalized uncertainty, whose Hilbert space representation~\cite{KempfManganoMann} leads to a position operator which has self-adjoint extensions defined on a set of continuous lattices, so that nearby points cannot be described by the same operator. In loop quantum gravity it is the area operator which is quantized~\cite{RovelliSmolin}, while an operatorial analysis of spacetime non commutativity in quantum field theory is in~\cite{BDFP}.

While we still lack a full theory of quantum gravity, it is nevertheless still possible to study field theories coupled with gravitational background, and gain fundamental insights of possible physics in that regime. 

In this letter we investigate the propagation of bosons. To this purpose we will use spectral techniques to study the actions. These techniques are ideally suited to tackle problems where the structure of spacetime may be fundamentally altered. The programme of noncommutative geometry~\cite{Connesbook} is in this direction, but the general ideas have a broader scope.  Finite mode regularization, based on the spectrum of the wave operator, was introduced in QCD~\cite{AndrianovBonora1, AndrianovBonora2, Fujikawabook}
The bosonic spectral action, appears not just in a context of noncommutative geometry, but also it naturally appears in QFT under the spectral regularization~\cite{Andrianov:2010nr,Andrianov:2011bc,Kurkov:2012dn}, for description of Weyl anomaly and also phenomena of induced Sakharov Gravity~\cite{Sakharov} and cosmological inflation\cite{Kurkov:2013gma}. 

For the scope of this paper we will use the spectral action, and the heat kernel techniques will be used to extract field theoretic information from it. The spectral action is defined in the presence of an energy scale $\Lambda$, which serves as cutoff. In this sense for us the high energy limit means in the proximity of $\Lambda$. 
Since the scale may signify a phase transition, what we are effectively investigating is the behaviour of these field as this phase transition occurs. The main result of this note is that the propagation of bosons effectively stops at high energy, in a precise way we describe below. We interpret this as an indication that the phase transition involves the fundaments of space time, and that at high scale points effectively decouple, giving rise to a ``pointless'' space. 

From the spectral point of view topological spaces are substituted by the equivalent concept of their algebras of continuous complex valued functions, while the geometry is encoded in a (generalized) Dirac operator which acts on a fermionic Hilbert space. The algebra of continuous functions is represented as operator on this same Hilbert space. In the ordinary cases the algebra is commutative, but the formalism is ready for the generalization to noncommutative spaces (see for example~\cite{WSS}). The spectral action~\cite{CCspectralaction,AC2M2} uses purely the spectral data of this generalized Dirac operator to describe, with an appropriate choice of the Dirac operator, the action of the usual standard model coupled with gravity. The model has some predictive power and can be confronted with experimental results, for recent results see~\cite{ColdPlay,CCvS} and references therein. Most of the calculations were done
with a few leading orders of the large $\Lambda$ asymptotic expansion of spectral action. Recently, it was shown
\cite{vanSuijlekom:2011kc} that including higher orders in the expansion leads to interesting consequences
for renormalization. The opposite (high momentum) asymptotics was considered in \cite{Iochum:2011yq}.

Here we are interested in the high energy propagators of bosons, and therefore we will use only the part of the Dirac operator which refers to spacetime, a generalization to the full operator of the standard model, or having any internal (gauge) degrees of freedom is immediate, and would just burden unnecessarily our notation. We shall work in Euclidean space and consider the high-momentum asymptotics, which we shall call the high energy asymptotics by slightly abusing the terminology.
Consider therefore the Euclidean Dirac operator\footnote{Here we consider a simplified Dirac operator, in particular we do not introduce the left-right fermion
doubling, necessary in the noncommutative geometry for the description of the standard model. The latter can be easily done, see Appendix 2, and will result just in an overall factor of two in all final formulas for high energy asymptotics.}
\begin{equation}
\mathcal{D}=\slashed{D}+\gamma_5 \phi \label{Dirop}
\end{equation}
where 
\begin{equation}
\slashed{D}=i\gamma^\mu \nabla_\mu = i\gamma^\mu (\nabla_\mu^{LC} + iA_\mu)\label{sD}
\end{equation}
is the usual geometric part with the Levi-Civita spin-connection and a gauge potential.
The spectral action associated to $\mathcal{D}$ is defined as
\begin{equation}
S_{f,\Lambda}\equiv {\rm Tr}\, f(\mathcal{D}^2/\Lambda) \,,\label{sact}
\end{equation}
where is a cut-off function restricted by the requirement that the trace in (\ref{sact}) exists,
and $\Lambda$ is a cut-off scale. Common choices for $f$ are a decreasing exponential, or the characteristic function of the unit interval, a sharp cutoff. 
On a non-compact space the spectral action (\ref{sact}) is 
divergent with the volume. To remove this divergence, it was suggested in~\cite{Iochum:2011yq}
to subtract the infrared divergence, i.e.\ replace  $f(\mathcal{D}^2/\Lambda)\to f(\mathcal{D}^2/\Lambda)-f(\mathcal{D}_0^2/\Lambda)$
under the trace. As $\mathcal{D}_0$ we take the free Dirac operator on the flat $\mathbb{R}^4$
with a zero gauge potential and a zero gauge field. Since the consequences of this subtraction are
rather obvious, we shall not write the $\mathcal{D}_0$-term explicitly in what follows.

For the choice $f(z)=e^{-z}$ the spectral action coincides with the heat trace
\begin{equation}
K(\mathcal{D}^2,s)={\rm Tr}\, \big( e^{-s\mathcal{D}^2}\big),\qquad s\equiv \Lambda^{-2}\,. \label{heatT}
\end{equation}
This will be our principal example in this work. As we shall argue later, our main results are valid
for generic cutoff function $f$ after suitable modifications.

Let us write $\mathcal{D}^2$ in the standard Laplace form \cite{Vassilevich:2003xt}
\begin{equation}
\mathcal{D}^2= -(\nabla^2+E)\,,\label{Lap}
\end{equation}
where
\begin{equation}
E=-i\gamma^\mu \gamma_5 (\phi_{,\mu}) -\phi^2 -\tfrac 14 R +\tfrac i4 [\gamma^\mu,\gamma^\nu]F_{\mu\nu} \label{Ephi}
\end{equation}
with $F_{\mu\nu}\equiv \partial_\mu A_\nu -\partial_\nu A_\mu$. The curvature of $\nabla$ reads:
\begin{equation}
\Omega_{\mu\nu}\equiv [\nabla_\mu,\nabla_\nu] = iF_{\mu\nu} -\tfrac 14 \gamma^\sigma \gamma^\rho 
R_{\sigma\rho\mu\nu}\,.\label{Omega}
\end{equation}

We would like to calculate the part of the spectral action/heat kernel that is responsible for propagation of bosons
at high momenta. This can be done with the help of the Barvinsky-Vilkovisky expansion~\cite{Barvinsky:1990up}
of the heat kernel for an operator $L$ of Laplace type (see Appendix 1). By taking the terms of this expansion
that are up to the quadratic order in the "curvatures" one can collect all terms that are quadratic in the fields.
The dependence on the momenta is expressed through the form-factors $f_1$ -- $f_5$, whose large-$\xi$ asymptotics
defines the large momenta asymptotics of the spectral action. To see how this works, let us consider the field
$\phi$ only. The relevant expression is in~\cite[Eq.~2.1]{Barvinsky:1990up} and is reproduced in Appendix 1 as Eq.~(\ref{BVex}). The term that depend on $\phi$ reads
\begin{equation}
K(L,s)_{(\phi)}\simeq (4\pi s)^{-2} \int d^4x\, {\rm tr}\,
\left[ sE + s^2 E \tfrac 12 h(-s\partial^2)E \right] \label{BVexp}
\end{equation}
With the help of (\ref{Ephi}), one finds at the order $\phi^2$:
\begin{equation}
K(\mathcal{D}^2,s)\simeq - \frac s{(4\pi s)^{-d/2}} \int  d^4x\, {\rm tr}\, 
\left[ \phi \left( 1 +\frac {s\partial^2}2 h(-s\partial^2) \right) \phi \right]\label{expphi}
\end{equation}
For large $(-s\partial^2)$:
\begin{equation}
K(\mathcal{D}^2,s)\simeq \frac {s}{(4\pi s)^{-d/2}} \int  d^4x\,  {\rm tr}\, \left[ \phi \frac 2{-s\partial^2} \phi \right]
\label{expphi2}
\end{equation}
Remarkably, the $\phi^2$-term is canceled, so that the heat kernel (and the spectral action) decay at large
momenta. The same effect was noted for gauge fields in \cite{Iochum:2011yq}, where also a cancellation of the
leading asymptotic expansion term took place. This is a quite interesting property of $\mathcal{D}^2$ as
compared to generic Laplace type operators.
 
Let us calculate the heat kernel to second order in metric perturbations over the flat background, 
$g_{\mu\nu}=\delta_{\mu\nu}+h_{\mu\nu}$.
We shall be interested in gravitons, i.e. in transverse traceless fluctuations: $\partial_\mu h_{\mu\nu}=0$,
$h_{\mu\mu}=0$. With the sign conventions of \cite{Vassilevich:2003xt}:
\begin{eqnarray}
&&\sqrt{g}_2= -\tfrac 14 h_{\mu\nu} h_{\mu\nu}\qquad 
\int  d^4x\,  (\sqrt{g}R)_2=\frac 14 \int d^4x\,  h_{\mu\nu} \partial^2 h_{\mu\nu}\\
&&(R_{\mu\nu\rho\sigma})_1=\frac 12 (\partial_\sigma \partial_\nu h_{\mu\rho}
+\partial_\mu\partial_\rho h_{\nu\sigma} 
-\partial_\mu\partial_\sigma h_{\nu\rho} 
-\partial_\nu\partial_\rho h_{\mu\sigma})\\
&&(R_{\nu\rho})_1=-\frac 12 \partial^2 h_{\nu\rho}\qquad (R)_1=0\,.
\end{eqnarray}
Here the subscript mean the order in the $h$. All indices are lower to stress that the summation over
repeated indices is performed with the Kronecker symbol rather than metric.
We can substitute these expansions in (\ref{BVex}) and calculate the
trace over spinor indices to obtain
\begin{equation}
K(\mathcal{D}^2,s)_{(h)}
\simeq \frac 1{(4\pi s)^2} \int  d^4x\,  h_{\mu\nu} \left[ -1 -\tfrac 1{12} s\partial^2 + (s\partial^2)^2
\left( f_1 (-s\partial^2) - \tfrac 12 f_5 (-s\partial^2) \right) \right] h_{\mu\nu} \label{Kgrav}
\end{equation}
Since there is no mixture between different fields, and quadratic terms in $A_\mu$ have already been calculated in \cite{Iochum:2011yq}, it just remains to pass to the $(-s\partial^2)\to\infty$ asymptotics and collect
everything together.  Finally, 
\begin{equation}
K(\mathcal{D}^2,s=1/\Lambda^2)\simeq  
\frac {\Lambda^4}{(4\pi )^2} \int d^4x \left[ -\tfrac 32  h_{\mu\nu}h_{\mu\nu}
+8 \phi \frac 1{-\partial^2} \phi + 8 F_{\mu\nu} \frac 1{(-\partial^2)^2} F_{\mu\nu} \right] \label{expan}
\end{equation}
Again, one notices the ``miraculous cancellation" of $h\partial^2 h$ and $F (1/\partial^2) F$ terms.

The spectral action for any other cut-off function $f$ can be obtained by an integral transformation of
the heat kernel. One can show, that due to this transformation the leading powers of the derivative
expansion remain the same, though numerical coefficients change (see \cite{Iochum:2011yq}, where the
gauge field expansion was analyzed in detail). Therefore, qualitatively our results remain valid for
a generic spectral action.

In order to interpret these results, and understand their physical meaning, we take the point of view that the cutoff is a physical scale up to which we may trust our theory, the natural candidate would be Planck's length. There is physical cutoff on length, which is imposed as a cutoff on the eigenvalues of the Dirac operator. This does not necessarily mean that there is a minimal length\footnote{For example the presence of $\Lambda_{\mathrm{QCD}}$ does not mean than in chromodynamics there is a maximal energy. There is however
a phase transition, related with confinement.}, although this is a possible interpretation. Our calculation indicates that rather than a minimal length, the cutoff indicates an energy in which the points of spacetime decouple.

We will see that a cutoff on the eigenvalues of the Dirac operator, and hence of the Laplacian, has profound consequences on the propagation of the fields. We are considering  free fields (i.e.\ plane waves), they are the ones one should use to probe spacetime. 
The propagator in position space $\Delta_F(x,y)$  has a meaning: the probability amplitude that a particle is created at position $x$, and later annihilated at position $y$. The probing of spacetime, in whichever scheme of realistic or gedanken experiment, involves always the interaction of particles, which are ``created'' in some apparatus, then interact with another particle at some position in space, and then are ``annihilated'' in a detector.

We assume homogeneity and isotropy, hence a two-point Green's functions depends on the difference between positions: $G(x-y)$. These are distributions acting on the space of test functions which physically are the sources $J(x)$. The latter are classical,  and we consider them to be the probes of spacetime. 
Let us now consider two situations, long and short distances. To probe short distances one requires high energetic sources. Mathematically this means that, in momentum space, the support of $J(k)$ is located in the large $k$ region. Using Eq.~\eqref{expan} it turns out, as we will discuss in more detail below, that asymptotically, in the high energy region, the Green's function becomes $\delta(x-y)$, or its derivatives. A source in $x$ has no effect on any other point, except $x$ itself. Heuristically, usually you have the vacuum, you ``disturb'' it with a source, and this disturbance propagates in a certain way, usually as a particle, generally a virtual one. Now instead we have that what happens in a point has no effect on neighbouring points. Points do not talk to each other.

Let us be more detailed. The classical action reads (in the quadratic field approximation):
\be
S[J,\phi]=\int d^4x \left(\frac12 \varphi(x) F(\del^2) \varphi(x) -J(x) \varphi(x)\right)\,,
\ee
where $\varphi$ is any of the bosonic fields, $\phi$, $A$, or $h$.
The equation of motion is:
\be
F(\del^2)\phi(x)=J(x)\label{eom}
\ee
The inverse operator
\be
G=\frac1{F(\del^2)}\ \ \ , \ \ G(k)=\frac1{F(-k^2)} 
\ee
allows us to write the solutions of (\ref{eom}) as
\be
\varphi_J(x)=\int d^4y J(y) G(x-y)=\frac1{(2\pi)^4}\int d^4k e^{ikx}J(k)\frac1{F(-k^2)}  \label{phix}
\ee
At low energy  $F(k)=k^2$, or $F(k)=k^2+m^2$, and everything is as we know. The Green's function is the usual one, leading to the normal propagation of particles.

The calculation above shows that in the very high energy regime (the scale is given by $\Lambda$) the qualitative behaviour has changed, and asymptotically $F(k)=1/k^2$ for scalars (and vectors), and $F(k)=1$ for gravitons. We now related this behaviour of $F$ with the nonpropagation, or better, to the impossibility to probe nearby points.

Short distances require high energetic probes, let us therefore consider $J(k)\neq 0$ for $|k^2|\in[K^2,K^2+\delta k^2]$, with $K^2$ very large. Substitute the expression for $F$ in~\eqn{phix} we obtain
\be
\varphi_J(x) \xrightarrow[{{\scriptstyle K\to\infty}}]{}
\left\{\begin{array}{ll} \frac1{(2\pi)^4}\int d^k e^{ikx} J(k) k^2=(-\del^2)J(x)  & \mbox{for scalars and vectors}  \\
\frac1{(2\pi)^4}\int d^k e^{ikx} J(k) =J(x)  & \mbox{for gravitons} 
\end{array}
\right. \label{phihigh}
\ee
What we find remarkable is the fact that the values of $\phi_j(x)$ depends only on $J$ or its derivatives calculated at $x$ itself. Compare with the standard case, in which to have the value at $x$ the whole function $J$ is required. In term of Green's function in position space, expression~\eqn{phihigh} means
\be
G(x-y)\propto\left\{\begin{array}{ll} (-\del^2)\delta\left(x-y\right)   & \mbox{for scalars and vectors}  \\
\delta\left(x-y\right)  & \mbox{for gravitons} 
\end{array}
\right. \label{green}
\ee
The Green's function vanish, unless $x=y$, hence there is no ``communication'' among points. 
There is no way test the topology, and know which point is near another point. At the mathematical level it can be shown~ that in the presence of a cutoff in the eigenvalues of the Dirac operator obtained with a projector (a sharp cutoff) the pure states of the algebra are at an infinite distance one from the other. In this case the distance is calculated using Connes formula, based on the Dirac operator.
This is the sort of spacetime one could expect from a transition in which the interaction among fields become infinitely strong, such as the one envisaged in~\cite{ULP}.

We are fully aware of the fact that we are stretching the field theory at its limit, and that we are in the realm where a full theory of quantum gravity should be employed. But failing this, we are using a theory we know as a pointer to a fully quantum gravitational phase.

There are still some unresolved question worth of investigation. Firstly is not clear if the cancellation of quadratic terms, which we have defined ``miraculous'' is a generic feature of the theory, and it has a deeper meaning related to the structure of the Dirac operator. 

The second issue is the physical value of $\Lambda$. On one side one would naturally assume that its natural value is the Planck mass. On the other side the fact that in the renormalization flow there are indications that some novelties should happen before. It has been known for a long time that the gauge coupling become nearly equal in a region of the order $10^{14}-10^{17}$~GeV (which is also the preferred range for $\Lambda$ in the spectral action approach to Standard Model). It was in fact believed for some time that that could signal the presence of a unification of the constants at a single point. 
Further experimental data have excluded this in the absence of new physics, although supersymmetric theory could still have a single unified point, for a recently updated review see~\cite[Sect.~6.4]{Martin:1997ns}. 
The presence of a relatively light Higgs particle at around 126~GeV may signal a phase of instability~\cite{instability} or metastability~\cite{metastability}. This may not just signal new physics, but it points at a new phase as well. The interpretation of $\Lambda$ at a lower scale than Planck mass opens the possibility of interesting new phenomena, especially from the cosmological point of view, in the intermediate regime between these two scales, in which a full theory of gravity as not yet fully set, but the probes of spacetime have started to behave in a way which points towards the regime described above.

\noindent\textbf{Acknowledgments:} DV was supported in parts by FAPESP, CNPq and by the INFN through
the Fondi FAI Guppo IV {\sl Mirella Russo} 2013. F.L. is partially supported by CUR Generalitat de Catalunya under project FPA2010-
20807.

\subsection*{Appendix 1: Barvinsky-Vilkovisky expansion}
In the paper \cite{Barvinsky:1990up} Barvinsky and Vilkovisky proposed an expansion of the heat kernel
to any (finite) order in $E$, $\Omega$ and the Riemann curvature, that is exact in the spectral
parameter $s$ and in $\partial^2$. To the leading orders this expansion reads
\begin{eqnarray}
K(L,s)&\simeq& \frac 1{(4\pi s)^2} \int d^4x g^{\frac 12}\, {\rm tr}\, \left[ 1 + sP + s^2\bigl( 
R_{\mu\nu} f_1(-s\partial^2) R^{\mu\nu} + Rf_2(-s\partial^2)R \right.\nn\\
&& \left. + Pf_3(-s\partial^2)R +Pf_4(-s\partial^2) P + \Omega_{\mu\nu} f_5(-s\partial^2)\Omega^{\mu\nu}
\bigr) \right] + \dots \label{BVex}
\end{eqnarray}
where $P\equiv E+\tfrac 16 R$,
\begin{eqnarray}
&&f_1(\xi)=\frac{h(\xi)-1+\tfrac 16 \xi}{\xi^2}\,,\qquad f_5(\xi)=-\frac {h(\xi)-1}{2\xi}\,,\nn\\
&&f_2(\xi)=\tfrac 1{288} h(\xi) -\tfrac 1{12} f_5(\xi) - \tfrac 18 f_1(\xi)\,,\qquad
f_3(\xi)=\tfrac 1{12} h(\xi) - f_5(\xi)\,,\nn\\
&&f_4=\tfrac 12 h(\xi)\,.
\end{eqnarray}
and
\begin{equation}
h(z):=\int_0^1 d\alpha \,e^{-\alpha (1-\alpha)\,z}\,. \label{h}
\end{equation}
Note, that the Barvinsky-Vilkovisky expansion is much more sensitive to the base space topology
than the standard heat kernel expansion. The formulas above are valid on $\mathbb{R}^4$ if all fields
vanish sufficiently fast at infinity, see \cite{Iochum:2012bu} for a more extended discussion. 

\subsection*{Appendix 2: Euclidean Dirac operator and fermion doubling}
We note, that the Bosonic Spectral action 
is defined for a Riemannian manifold with Euclidean signature of metric.
In contrast to the bosonic case, the ``Euclidisation" of
fermions is not just analytical continuation but is more delicate issue.
One way of the Euclidisation, being the most suitable for the noncommutative geometry (see~\cite{doubling,GraciaBondia:1997za,Schucker:2001aa} for discussions),
 is based on the doubling of fermionic
degrees of freedom.
The idea is the following: each two component chiral spinor of the SM must be replaced by the four component
Dirac fermion, and left and right fermions are treated as independent degrees of freedom,
in particular
\be
\psi^{\rm Eucl}_{\rm L} \neq \frac{1}{2}\left(1 - \gamma_5 \right)\psi^{\rm Eucl}_{\rm L}, 
\quad \psi^{\rm Eucl}_{\rm R} \neq \frac{1}{2}\left(1 + \gamma_5 \right)\psi^{\rm Eucl}_{\rm R}.
\ee
We stress, that {\it both} $\psi^{\rm Eucl}_{\rm L}$ and $\psi^{\rm Eucl}_{\rm R}$ have
four independent components each, i.e. 8 independent components totally.
It is important, that when one computes the partition function $Z$ or conformal anomaly, RG equations and etc. one must put {\it by hand}
a factor of $1/2$, where needed, e.g. $Z^{\rm Mink} = \left(Z^{\rm Eucl}\right)^{\frac{1}{2}}$. 
{\it Only} when one comes back to Minkowski signature one reduces number of fermions, imposing the projection 
\be
\psi^{\rm Mink}_{\rm L} = \frac{1}{2}\left(1 - \gamma_5 \right)\psi^{\rm Mink}, 
\quad \psi^{\rm Mink}_{\rm R} = \frac{1}{2}\left(1 + \gamma_5 \right)\psi^{\rm Mink}.
\ee 

The Hilbert space has the following structure:
\be
H = H_{\rm L} \oplus H_{\rm R},
\ee
where $H_{\rm L}$ and $H_{\rm R}$ are spaces
of left and right (four component) fermions, and the Higgs field $\phi$ {\it connects} 
left and right fermions. In case of a single massive fermion, the classical action reads (c.f. \cite[Appendix]{CCspectralaction}),
\bea
S_{\rm F} &=& \int d^4 x\,\sqrt{g}\, \Psi^{\dagger} \mathcal{D} \Psi,\no
\Psi &\equiv&   \left(\begin{array}{l} \psi_{\rm L} \\
        \psi_{\rm R}\end{array} \right),\no
   \mathcal{D} &\equiv& i\gamma^\mu (\nabla_\mu^{LC} + iA_\mu)\otimes 1_2^{\rm{L-R}}      + 
    \gamma_5\, \phi\, \otimes \sigma_2^{\rm L-R}, \la{DiracN}
\eea
where 
\be
1_{2}^{\rm L-R} =   \left(\begin{array}{ll}
         1 & 0 \\
        0 & 1\end{array} \right) \quad,
        \sigma_{1}^{\rm L-R} =   \left(\begin{array}{ll}
         0 & 1 \\
        1 & 0 \end{array} \right)
\ee   
are matrices, acting on $\rm L$ and $\rm R$ indices. 



\end{document}